\documentclass[smallabstract,smallcaptions]{dccpaper}

\usepackage{citesort}
\usepackage{amsmath,amssymb}
\usepackage{url}
\usepackage{array,multirow,booktabs}

\newlength{\figurewidth}
\newlength{\smallfigurewidth}
\usepackage{bm}
\def\V{{\bm V}}
\def\U{{\bm U}}
\def\btheta{{\bm \theta}}
\def\htheta{{\hat \btheta}}
\def\I{{\bm I}}
\def\hV{{\hat{\V}}}

\def\etal{{\textit{et al}.}}

\usepackage{tikz}
\usetikzlibrary{backgrounds,calc,chains,fit,matrix,positioning, arrows.meta,shadows,shapes.misc,circuits.ee,shapes}
\tikzset{input/.style={}}
\tikzset{output/.style={}}
\tikzset{operator/.style={circle, draw, fill=black!8, minimum size=2.5ex, inner sep=0pt}}
\tikzset{filter/.style={rectangle, draw, fill=black!8, minimum size=3.5ex, inner xsep=1.5ex, align=center}}
\tikzset{filter1/.style={rectangle, draw, dashed, fill=black!8, minimum size=3.5ex, inner xsep=1.5ex, align=center}}
\tikzset{other/.style={rounded rectangle, draw, fill=black!8, minimum size=3.5ex, inner xsep=1ex}}
\tikzset{branch/.style={circle, draw, fill=black, minimum size=.5ex, inner sep=0pt}}
\tikzset{rv/.style={circle, draw, thick, fill=white, minimum size=2.75ex, inner sep=0pt}}
\tikzset{ob/.style={circle, draw, thick, fill=lightgray, minimum size=2.75ex, inner sep=0pt}}
\tikzset{pa/.style={circle, draw, thick, fill=black, minimum size=1ex, inner sep=0pt}}
\tikzset{/tikz/thin/.style={line width=.9pt}}
\tikzset{/tikz/thick/.style={line width=1.4pt}}
\tikzset{every path/.style={thin}}
\tikzset{>=direction ee}
\tikzset{every loop/.style={min distance=10mm,in=-60,out=60,looseness=10}} %
\usepackage{pgfplots}
\pgfplotsset{compat=1.14}
\pgfplotsset{every axis/.append style={enlargelimits={abs=3pt},grid,axis lines=left}}
\pgfplotsset{every axis plot/.append style={thick,mark size=1.5pt,line join=bevel,mark options={solid}}}
\pgfplotsset{label style={font=\small}}
\pgfplotsset{tick label style={font=\footnotesize}}
\pgfplotsset{grid style={color=black!10}}
\pgfplotsset{legend style={draw=none,opacity=.85,font=\footnotesize,cells={anchor=west,opacity=1}}}
\pgfplotsset{every non boxed x axis/.style={xtick align=center,shorten <=-.5\pgflinewidth}}
\pgfplotsset{every non boxed y axis/.style={ytick align=center,shorten <=-.5\pgflinewidth}}
\pgfplotsset{every non boxed z axis/.style={ztick align=center,shorten <=-.5\pgflinewidth}}
\pgfplotsset{/pgf/number format/1000 sep={\,}}

\newcommand{\hpsrpccDecoder}{
\begin{tikzpicture}[
    every label/.style={align=center},
    myfont/.style={
    },
    cell/.style={
        circle,draw, fill=black!8,
        line width = .75pt,
        minimum width=1cm,
        inner sep=1pt,
    },
    corner/.style={
        circle,draw, fill=black!8,
        line width = .75pt,
        inner sep=1pt,
    },
    bitstream/.style={
        ellipse,
        draw, very thick, fill=black!8,
        minimum width=1cm,
        minimum height=0.66cm,
        inner sep=1pt,
    },
    hpsr/.style={
        rectangle, draw, fill=black!8,
        rounded corners=5mm,
        minimum width=4cm,
        minimum height=2cm,
        inner sep=1pt,
        align=center, 
        very thick,
    },
    coder/.style={
        rectangle, draw, fill=black!8,
        minimum width=1.4cm,
        minimum height=1cm,
        inner sep=1pt,
        align=center, %
        very thick,
    },
    ArrowC1/.style={
        rounded corners=2mm,
        >=Stealth[round],
        very thick,
    },
]
    \node [hpsr] (hpsr) at (2.5,0) { \textbf{Iterative Inference with} \\ \textbf{Super Resolution Network}\\ $f(\cdot;\htheta)$};
    \node[cell, label={[myfont]above:Reconstructed\\ Point Cloud}] (rpc) at (-1.8,0) {$\hV$};
    \node[cell, label={[myfont]above:Base\\ Point Cloud}] (bpc) at (7,0) {${\V}^{(K)}$};
    \node[cell, label={[myfont]below:Reconstructed\\ Network Parameter}] (hp) at (7,-2) {$\htheta$};
    \node [coder] (baseCoder) at (10,0) {Base \\ Decoder};
    \node [coder] (sideCoder) at (10,-2) {Side \\ Decoder};
    \node[bitstream, label={[myfont]above:Base\\ Bitstream}] (basebit) at (12,0) {};
    \node[bitstream, label={[myfont]below:Side\\ Bitstream}] (sidebit) at (12,-2) {};
    \draw [->, ArrowC1] (basebit) -- (baseCoder) -- (bpc) -- (hpsr);
    \draw [ArrowC1] (sidebit) -- (sideCoder) -- (hp);
    \draw [->, ArrowC1] (hp) -| (hpsr);
    \draw [->, ArrowC1] (hpsr) -- (rpc);   
\end{tikzpicture}
}
\newcommand{\hpsrpccEncoder}{
\begin{tikzpicture}[
    every label/.style={align=center},
    myfont/.style={
    },
    frame/.style={
        rectangle,
        rounded corners=5mm,
        draw,gray, fill=black!8,
        very thick,
    },
    cell/.style={
        circle,draw, fill=black!8,
        line width = .75pt,
        minimum width=1cm,
        inner sep=1pt,
    },
    corner/.style={
        circle,draw, fill=black!8,
        line width = .75pt,
        inner sep=1pt,
    },
    bitstream/.style={
        ellipse,
        draw,very thick, fill=black!8,
        minimum width=1cm,
        minimum height=0.66cm,
        inner sep=1pt,
    },
    downfunc/.style={
        rectangle, draw, fill=black!8,
        minimum width=1cm,
        minimum height=1cm,
        inner sep=1pt,
        very thick,
    },
    coder/.style={
        rectangle, draw, fill=black!8,
        minimum width=1.4cm,
        minimum height=1cm,
        inner sep=1pt,
        align=center, 
        very thick,
    },
    hpc/.style={
        rectangle, fill=black!8,
        rounded corners=5mm,
        draw,
        minimum width=4.5cm,
        minimum height=2cm,
        inner sep=1pt,
        align=center,
        very thick,
    },
    ArrowC1/.style={
        rounded corners=3mm,
        >=Stealth[round],
        very thick,
    },
]
    \node [frame, minimum height =3cm, minimum width=6cm, label={above:\textbf{Two-Stage Downsampling}}] at (0,0) {};
    \node [hpc] (hpc) at (2.8,-3) {\textbf{Super Resolution Network} \\ $f(\cdot;\btheta)$ \textbf{Optimization}};
    
    \node [downfunc] (down0) at (-2,0) {$\downarrow 2^{1-K}$};
    \node[cell, label={[myfont]above:Intermediate\\ Point Cloud}] (ipc) at (0,0) {${\U}^{(K)}$};
    \node [downfunc] (downK) at (2,0) {$\downarrow 2^{-1}$};

    \node[cell, label={[myfont]above:Original\\ Point Cloud}] (pc) at (-4.6,0) {{$\V$}};

    \node[cell, label={[myfont]above:Base\\ Point Cloud}] (bpc) at (6,0) {${\V}^{(K)}$};
    \node[corner] (corner) at (-2,-1) {};
    \node[cell, label={[myfont]below:Point Cloud Pair}] (pcp) at (-2,-3) {${\U}^{(K)}, {\V}^{(K)}$};
    \node[cell, label={[myfont]below:Network\\ Parameters}] (hp) at (6.8,-3) {$\btheta$};

    \node [coder] (baseCoder) at (10,0) {Base \\ Encoder};
    \node [coder] (sideCoder) at (10,-3) {Side \\ Encoder};
    
    \node[bitstream, label={[myfont]above:Base\\ Bitstream}] (basebit) at (12,0) {};
    \node[bitstream, label={[myfont]below:Side\\ Bitstream}] (sidebit) at (12,-3) {};

    \draw [->, ArrowC1] (pc) -- (down0) -- (ipc) -- (downK) -- (bpc); 
    \draw [->, ArrowC1] (bpc) -- (baseCoder) -- (basebit);
    \draw [->, ArrowC1] (down0) -- (pcp);
    \draw [ArrowC1] (downK) |- (corner);
    \draw [->, ArrowC1] (pcp) -- (hpc) -- (hp) -- (sideCoder) -- (sidebit);

\end{tikzpicture}
}

\newcommand{\casea}{
\begin{tikzpicture}
  \draw[yslant=-0.5]  (0,0) rectangle +(2,2);
  \draw[yslant=-0.5] (0,0) grid (2,2);
  
  \draw[yslant=0.5] (2,-2) rectangle +(2,2);
  \draw[yslant=0.5] (2,-2) grid (4,0);
  \fill[yslant=0.5, gray!80] (3,-1) rectangle +(1,1);
  
  \draw[yslant=0.5,xslant=-1,] (4,2) rectangle +(-2,-2);
  \draw[yslant=0.5,xslant=-1] (2,0) grid (4,2);
  \fill[yslant=0.5,,xslant=-1, gray!80] (3,0) rectangle +(1,1);

  \draw[-latex, line width=2pt] (4,1) -- (9.5,1) node[midway] (quantization) {};
  \node [above=2pt of quantization] {\Huge downsampling};

  \fill[yslant=-0.5, gray!80] (9.5,5) rectangle +(2,2);
  \fill[yslant=0.5, gray!80] (11.5,-6.5) rectangle +(2,2);
  \fill[yslant=0.5,,xslant=-1, gray!80] (9,-2.5) rectangle +(-2,-2);
  
  \draw[-latex, line width=2pt] (13.5,1) -- (17,1) node[midway] (inverse quantization) {};
  \node [above=2pt of inverse quantization] {\Huge upscaling};

  \draw[yslant=-0.5]  (17,9) rectangle +(2,2);
  \draw[yslant=-0.5] (17,9) grid +(2,2);
  
  \draw[yslant=0.5] (19,-10) rectangle +(2,2);
  \draw[yslant=0.5] (19,-10) grid +(2,2);
  \fill[yslant=0.5, gray!80] (20,-9) rectangle +(1,1);
  
  \draw[yslant=0.5,xslant=-1,] (13,-6) rectangle +(-2,-2);
  \draw[yslant=0.5,xslant=-1] (13,-6) grid +(-2,-2);
  \fill[yslant=0.5,,xslant=-1, gray!80] (12,-8) rectangle +(1,1);
\end{tikzpicture}
}

\newcommand{\caseb}{
\begin{tikzpicture}
  \draw[yslant=-0.5]  (0,0) rectangle +(2,2);
  \draw[yslant=-0.5] (0,0) grid (2,2);
  \fill[yslant=-0.5, gray!80] (0,1) rectangle +(1,1);
  \fill[yslant=-0.5, gray!80] (1,0) rectangle +(1,1);
  
  \draw[yslant=0.5] (2,-2) rectangle +(2,2);
  \draw[yslant=0.5] (2,-2) grid (4,0);
  \fill[yslant=0.5, gray!80] (2,-2) rectangle +(1,1);
  \fill[yslant=0.5, gray!80] (3,-1) rectangle +(1,1);
  
  \draw[yslant=0.5,xslant=-1,] (4,2) rectangle +(-2,-2);
  \draw[yslant=0.5,xslant=-1] (2,0) grid (4,2);
  \fill[yslant=0.5,,xslant=-1, gray!80] (2,1) rectangle +(1,1);
  \fill[yslant=0.5,,xslant=-1, gray!80] (3,0) rectangle +(1,1);

  \draw[-latex, line width=2pt] (4,1) -- (9.5,1) node[midway] (quantization) {};
  \node [above=2pt of quantization] {\Huge downsampling};

  \fill[yslant=-0.5, gray!80] (9.5,5) rectangle +(2,2);
  \fill[yslant=0.5, gray!80] (11.5,-6.5) rectangle +(2,2);
  \fill[yslant=0.5,,xslant=-1, gray!80] (9,-2.5) rectangle +(-2,-2);
  
  \draw[-latex, line width=2pt] (13.5,1) -- (17,1) node[midway] (inverse quantization) {};
  \node [above=2pt of inverse quantization] {\Huge upscaling};

  \draw[yslant=-0.5]  (17,9) rectangle +(2,2);
  \draw[yslant=-0.5] (17,9) grid +(2,2);
  \draw[yslant=-0.5, black, line width=4pt] (17,10) rectangle +(1,1);
  \draw[yslant=-0.5, black, line width=4pt] (18,9) rectangle +(1,1);
  
  \draw[yslant=0.5] (19,-10) rectangle +(2,2);
  \draw[yslant=0.5] (19,-10) grid +(2,2);
  \fill[yslant=0.5, gray!80] (20,-9) rectangle +(1,1);
  \draw[yslant=0.5, black, line width=4pt] (19,-10) rectangle +(1,1);
  
  \draw[yslant=0.5,xslant=-1,] (13,-6) rectangle +(-2,-2);
  \draw[yslant=0.5,xslant=-1] (13,-6) grid +(-2,-2);
  \fill[yslant=0.5,,xslant=-1, gray!80] (12,-8) rectangle +(1,1);
  \draw[yslant=0.5,,xslant=-1, black, line width=4pt] (11,-7) rectangle +(1,1);
\end{tikzpicture}
}

\newcommand{\casec}{

\begin{tikzpicture}
  \draw[yslant=-0.5]  (0,0) rectangle +(2,2);
  \draw[yslant=-0.5] (0,0) grid (2,2);
  \fill[yslant=-0.5, gray!80] (0,1) rectangle +(1,1);
  
  \draw[yslant=0.5] (2,-2) rectangle +(2,2);
  \draw[yslant=0.5] (2,-2) grid (4,0);
  
  \draw[yslant=0.5,xslant=-1,] (4,2) rectangle +(-2,-2);
  \draw[yslant=0.5,xslant=-1] (2,0) grid (4,2);
  \fill[yslant=0.5,,xslant=-1, gray!80] (2,1) rectangle +(1,1);

  \draw[-latex, line width=2pt] (4,1) -- (9.5,1) node[midway] (quantization) {};
  \node [above=2pt of quantization] {\Huge downsampling};

  \fill[yslant=-0.5, gray!80] (9.5,5) rectangle +(2,2);
  \fill[yslant=0.5, gray!80] (11.5,-6.5) rectangle +(2,2);
  \fill[yslant=0.5,,xslant=-1, gray!80] (9,-2.5) rectangle +(-2,-2);
  
  \draw[-latex, line width=2pt] (13.5,1) -- (17,1) node[midway] (inverse quantization) {};
  \node [above=2pt of inverse quantization] {\Huge upscaling};

  \draw[yslant=-0.5]  (17,9) rectangle +(2,2);
  \draw[yslant=-0.5] (17,9) grid +(2,2);
  \draw[yslant=-0.5, black, line width=4pt] (17,10) rectangle +(1,1);
  
  \draw[yslant=0.5] (19,-10) rectangle +(2,2);
  \draw[yslant=0.5] (19,-10) grid +(2,2);
  \fill[yslant=0.5, black] (20,-9) rectangle +(1,1);
  
  \draw[yslant=0.5,xslant=-1,] (13,-6) rectangle +(-2,-2);
  \draw[yslant=0.5,xslant=-1] (13,-6) grid +(-2,-2);
  \fill[yslant=0.5,,xslant=-1, black] (12,-8) rectangle +(1,1);
  \draw[yslant=0.5,,xslant=-1, black, line width=4pt] (11,-7) rectangle +(1,1);
\end{tikzpicture}
}

\newcommand{\cased}{
\begin{tikzpicture}
  \draw[yslant=-0.5]  (0,0) rectangle +(2,2);
  \draw[yslant=-0.5] (0,0) grid (2,2);
  \fill[yslant=-0.5, gray!80] (0,1) rectangle +(1,1);
  \fill[yslant=-0.5, gray!80] (1,0) rectangle +(1,1);
  
  \draw[yslant=0.5] (2,-2) rectangle +(2,2);
  \draw[yslant=0.5] (2,-2) grid (4,0);
  \fill[yslant=0.5, gray!80] (2,-2) rectangle +(1,1);
  
  \draw[yslant=0.5,xslant=-1,] (4,2) rectangle +(-2,-2);
  \draw[yslant=0.5,xslant=-1] (2,0) grid (4,2);
  \fill[yslant=0.5,,xslant=-1, gray!80] (2,1) rectangle +(1,1);

  \draw[-latex, line width=2pt] (4,1) -- (9.5,1) node[midway] (quantization) {};
  \node [above=2pt of quantization] {\Huge downsampling};

  \fill[yslant=-0.5, gray!80] (9.5,5) rectangle +(2,2);
  \fill[yslant=0.5, gray!80] (11.5,-6.5) rectangle +(2,2);
  \fill[yslant=0.5,,xslant=-1, gray!80] (9,-2.5) rectangle +(-2,-2);
  
  \draw[-latex, line width=2pt] (13.5,1) -- (17,1) node[midway] (inverse quantization) {};
  \node [above=2pt of inverse quantization] {\Huge upscaling};

  \draw[yslant=-0.5]  (17,9) rectangle +(2,2);
  \draw[yslant=-0.5] (17,9) grid +(2,2);
  \draw[yslant=-0.5, black, line width=4pt] (17,10) rectangle +(1,1);
  \draw[yslant=-0.5, black, line width=4pt] (18,9) rectangle +(1,1);
  
  \draw[yslant=0.5] (19,-10) rectangle +(2,2);
  \draw[yslant=0.5] (19,-10) grid +(2,2);
  \fill[yslant=0.5, black] (20,-9) rectangle +(1,1);
  \draw[yslant=0.5, black, line width=4pt] (19,-10) rectangle +(1,1);
  
  \draw[yslant=0.5,xslant=-1,] (13,-6) rectangle +(-2,-2);
  \draw[yslant=0.5,xslant=-1] (13,-6) grid +(-2,-2);
  \fill[yslant=0.5,,xslant=-1, black] (12,-8) rectangle +(1,1);
  \draw[yslant=0.5,,xslant=-1, black, line width=4pt] (11,-7) rectangle +(1,1);
  
\end{tikzpicture}

}

\begin{document}

\title
{\large
\textbf{Lightweight super resolution network for \\ point cloud geometry compression}\thanks{The first two authors contribute equally. This work was supported by The Major Key Project of PCL. Corresponding author: Wen Gao.}
}

\author{%
Wei Zhang$^{\star\ast}$, Dingquan Li$^{\star}$, Ge Li$^{\dagger}$, and Wen Gao$^{\star\ast}$\\[0.5em]
{\small\begin{minipage}{\linewidth}\begin{center}
\begin{tabular}{c}
$^{\star}$Peng Cheng Laboratory, Shenzhen, China \\
$^{\ast}$Harbin Institute of Technology, Shenzhen, China \\ 
$^{\dagger}$Peking University Shenzhen Graduate School, Shenzhen, China \\
\end{tabular}
\end{center}\end{minipage}}
}

\maketitle
\thispagestyle{empty}

\begin{abstract}
This paper presents an approach for compressing point cloud geometry by leveraging a lightweight super-resolution network. The proposed method involves decomposing a point cloud into a base point cloud and the interpolation patterns for reconstructing the original point cloud. While the base point cloud can be efficiently compressed using any lossless codec, such as Geometry-based Point Cloud Compression, a distinct strategy is employed for handling the interpolation patterns. Rather than directly compressing the interpolation patterns, a lightweight super-resolution network is utilized to learn this information through overfitting. Subsequently, the network parameter is transmitted to assist in point cloud reconstruction at the decoder side. Notably, our approach differentiates itself from lookup table-based methods, allowing us to obtain more accurate interpolation patterns by accessing a broader range of neighboring voxels at an acceptable computational cost. Experiments on MPEG Cat1 (Solid) and Cat2 datasets demonstrate the remarkable compression performance achieved by our method. 
\end{abstract}

\Section{Introduction}
Point clouds have gained significant popularity due to their remarkable ability to realistically and naturally represent 3D objects. They serve as a versatile media format widely used in numerous immersive multimedia applications, including augmented, virtual, and mixed reality~\cite{schwarz2019emerging}. Point cloud compression (PCC) is a rapidly growing research field focused on creating efficient coding solutions for increasingly large point clouds to meet storage and transmission requirements. Leading standardization bodies, such as the Moving Picture Experts Group (MPEG), the Joint Photographic Experts Group (JPEG), and the Audio Video Coding Standards Workgroup of China (AVS), have dedicated substantial efforts to establish PCC standards~\cite{tmc2v22,tmc13v22,mpeg-aipcc-cfp,jpeg-pcc2023vm,avspcc}. Concurrently, there has been a surge of interest in scientific literature, with various researchers working on improved compression methods for point clouds~\cite{quach2022survey,cao2021compression,li2023hierarchical,borges2022fractional,wang2023sparse,fu2022octattention}. In this context, we focus on point cloud geometry compression, and we review related and representative works in this field below.

Video-based PCC (V-PCC) and Geometry-based PCC (G-PCC) represent two conventional approaches~\cite{graziosi2020overview}. V-PCC employs 3D-to-2D projections, leveraging well-established video coding techniques for compression. Over time, several enhancements have been introduced to V-PCC, addressing aspects such as rate control, plane projection, reduction of computational complexity, motion prediction, and artifact removal.
In contrast, G-PCC operates directly on 3D point clouds, offering the potential for superior compression performance by fully exploiting the inherent data correlations. G-PCC (octree) represents point clouds using an octree structure and encodes the octree using context-based arithmetic coding. G-PCC (trisoup) represents 3D cubes using surfaces composed of triangle strips and encodes the triangles instead of the individual points within these cubes. To enhance the quality of decompressed point clouds in G-PCC, a lookup table-based post-processing super-resolution method has been proposed~\cite{borges2022fractional}. Li~\etal~\cite{li2023hierarchical} have developed a more accurate lookup table at the encoder side and transmitted it to the decoder to achieve improved super resolution and better rate-distortion optimization. However, it is important to note that the overhead associated with transmitting the lookup table increases exponentially concerning the considered neighboring voxels, even though it leads to reduced distortion. To address this issue, we propose an alternative approach. Instead of transmitting the lookup table, we employ a lightweight super resolution network to approximate it, resulting in a significantly reduced overhead for transmitting the network parameter and a more efficient compression scheme.

There is a growing recognition of the competitiveness of learning-based PCC solutions~\cite{quach2022survey}. Based on the representation form of point clouds, these methods can be broadly classified into three categories: voxel-based, octree-based, and point-based PCC. Voxel-based methods treat the point cloud geometry as a binary signal on a voxel grid and employ 3D (sparse) convolutions~\cite{nguyen2021learning,wang2023sparse}. Octree-based methods represent the point cloud geometry using an octree structure and predict the children of the current node based on its neighboring nodes, such as ancestors and siblings~\cite{huang2020octsqueeze,fu2022octattention,chen2022point}. Point-based methods utilize PointNet-like architectures to learn a compact representation for point clouds~\cite{huang20193d,wen2020lossy}. It is worth noting that all of these methods train their models on extensive point cloud datasets, and these models are neither lightweight nor tailored to the specific content of the point cloud being compressed.

In contrast, this paper introduces a lightweight super resolution network for content-adaptive point cloud geometry compression. Specifically, we decompose a point cloud into a base point cloud and its interpolation patterns, which are essential for reconstructing the original point cloud. This approach enables us to achieve compression for the original point cloud by effectively compressing both the base point cloud and its interpolation patterns. For the compression of the base point cloud, we employ a pre-existing lossless codec, such as G-PCC or OctAttention~\cite{fu2022octattention}, to ensure efficient compression. To handle the interpolation patterns, we present an alternative strategy. Instead of directly compressing this information, we opt for overfitting a lightweight super resolution network to learn it. Subsequently, we encode and transmit the content-adaptive network parameter, which facilitates point cloud reconstruction at the decoder side. Our method is validated through experimental verification using the MPEG Cat1 (Solid) and Cat2 datasets. The results of these experiments underscore the effectiveness of our approach, demonstrating significant Bj{\o}ntegaard-delta bitrate savings compared to  
G-PCC (octree) v22 and V-PCC v22.

\Section{Background: Downsampling and Upscaling}\label{sec:modelling}
We represent the original point cloud as $\V=\left\{(x_i,y_i,z_i)\right\}_{i=1}^N$ in a voxelized form, where $N$ denotes the number of points. By applying a downsampling operation with a factor of $2$ to $\V$, we obtain a downsampled point cloud, $\V_d=\left\{({x_d}_j,{y_d}_j,{z_d}_j)\right\}_{j=1}^M$, denoted as follows:
\begin{equation}
    \V_d = \mathrm{unique}\left(\left[\V/2\right]\right).
\end{equation}
Here, the function $\mathrm{unique}$ removes any duplicate points, and $\left[\cdot\right]$ represents the rounding function. 

\begin{figure}[!t]
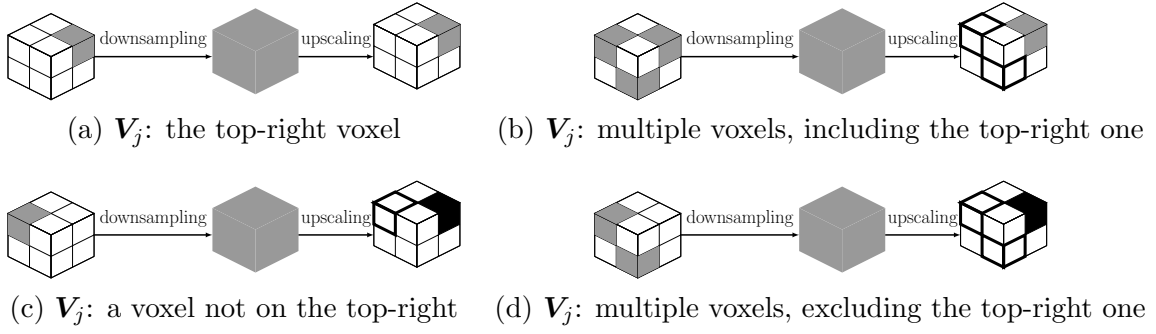

\begin{center}
\begin{tabular}{cc}
\resizebox{.4\linewidth}{!}{\casea} & \resizebox{.4\linewidth}{!}{\caseb} \\
{\small (a) $\V_j$: the top-right voxel} & {\small (b) $\V_j$: multiple voxels, including the top-right one} \\[1em]
\resizebox{.4\linewidth}{!}{\casec}
 & \resizebox{.4\linewidth}{!}{\cased} \\
{\small (c) $\V_j$: a voxel not on the top-right} & {\small (d) $\V_j$: multiple voxels, excluding the top-right one}
\end{tabular}
\end{center}
\caption{Illustration of possible scenarios during the downsampling and upscaling processes. In each subfigure, the left side depicts the voxelization of a subset $\V_j$, where the gray-filled cubes represent the presence of points (occupied voxels), and the white-unfilled cubes signify the absence of points  (empty voxels). In the middle, following a $2\times$ downsampling, the point $({x_d}_j, {y_d}_j,{z_d}_j)$ in the downsampled point cloud $\V_d$ is obtained. On the right, upscaling results in the reconstruction of $\hV_j$, which includes only a single point $(2{x_d}_j, 2{y_d}_j,2{z_d}_j)$. The black-filled cubes represent newly added incorrect points, while the unfilled cubes with black lines indicate the failure to recover the original points. }\label{fig:quantization-and-inverse} 
\end{figure}

Based on the downsampled point cloud $\V_d$, the original point cloud $\V$ can be partitioned into $M$ subsets $\left\{\V_j\right\}_{j=1}^{M}$. Each subset, $\V_j$, encompasses all the points in $\V$ that map to $({x_d}_j,{y_d}_j,{z_d}_j)$ after downsampling, which can be expressed as follows:
\begin{align}
    \V = & \bigcup\limits_{j=1}^{M} \V_j,\\ 
    \V_j = & \left\{(x_i,y_i,z_i)\in\V\mid\left[\frac{x_i}{2}\right]={x_d}_j, \left[\frac{y_i}{2}\right]={y_d}_j, \left[\frac{z_i}{2}\right]={z_d}_j\right\}.
\end{align}
In accordance with this definition, each subset $V_j$ comprises a minimum of $1$ element and a maximum of $8$ elements. Fig.~\ref{fig:quantization-and-inverse} illustrates possible scenarios of downsampling and upscaling processes. Scenario (a) does not induce distortion and allows for the complete recovery of information in  $\V_j$. In all other scenarios, distortions arise due to the inclusion of incorrect points and/or the loss of original points. 
Only when all $\left\{\V_j\right\}_{j=1}^{M}$ adhere to Scenario (a), encoding $\V_d$ losslessly  preserves information about $\V$, achieving lossless compression of $\V$.

\Section{Proposed Method}
There exists a one-to-one or one-to-many mapping between the downsampled point $({x_d}_j,{y_d}_j,{z_d}_j)$ and its corresponding original points in $\V_j$ for $j=1,\cdots,M$. We refer to this information as \textit{interpolation patterns}, denoted as $\I_u$. If we encode these interpolation patterns, we can achieve point cloud super resolution using them rather than direct upscaling, which serves to reduce or even remove the aforementioned distortions. In essence, by decomposing the original point cloud into a downsampled point cloud and its corresponding interpolation patterns, the problem of lossy/lossless compression of the original point cloud is transformed into lossless compression of the downsampled point cloud and lossy/lossless compression of the interpolation patterns. This approach resembles the ``base layer + enhancement layer'' framework commonly used in video coding, such as LCEVC~\cite{battista2022overview}. The downsampled point cloud can be effectively compressed using any lossless codec. The key of our approach hinges on the compression of the interpolation patterns. Directly compressing the interpolation patterns may incur an enormous bitrate cost. Therefore, we propose employing a neural network to predict the interpolation patterns $\I_u$ from the downsampled point cloud $\V_d$. Subsequently, we encode and transmit the network parameter instead.

\SubSection{Prediction of the Interpolation Pattern}
Let ${\bm u}_j^{4\delta_x+2\delta_y+\delta_z}=(2{x_d}_j-\delta_x, 2{y_d}_j-\delta_y,2{z_d}_j-\delta_z)$, where $\delta_x, \delta_y, \delta_z \in \{0,1\}$. We obtain the interpolation pattern corresponding to the point $({x_d}_j, {y_d}_j,{z_d}_j)$ in $\V_d$ by determining whether the point ${\bm u}_j^k$ belongs to the subset $\V_j$:
\begin{equation}
    {I_u}(j,k) = \mathbb{I}\left[{\bm u}_j^k\in \V_j\right], \quad k=0,\cdots, 7, \quad j=1, \cdots, M.
\end{equation}
Here, $\mathbb{I}\left[\cdot\right]$ is the indicator function, taking the value of $1$ when the event inside the brackets is true and $0$ otherwise. Consequently, the original point cloud can be transformed into the representation of the downsampled point cloud $\V_d$ and its corresponding interpolation patterns $\I_u$:
\begin{equation}
    \V\longrightarrow\left\{\V_d, \I_u\right\}.
\end{equation}

In theory, given that the downsampled point cloud $\V_d$ will be losslessly encoded, we can input all the information of $\V_d$ into a neural network $f(\cdot;{\btheta})$ to predict $\I_u$:
\begin{equation}\label{eq:ori_model}
    \I_u = f(\V_d;{\btheta}).
\end{equation}

However, including excessive reference information would increase the complexity of the neural network, which is not conducive to the compression of the neural network. We observe that point clouds exhibit non-local geometric similarities, implying that when the occupancy information of neighboring points in the downsampled point cloud is the same, the interpolation pattern will likely be the same. Let $\mathcal{N}_j$ denote the set of neighbors of the point $({x_d}_j, {y_d}_j,{z_d}_j)$ in $V_d$. The occupancy information can be represented as follows:
\begin{equation}
    {\bm O}_{\mathcal{N}_j} = \left\{\mathbb{I}\left[{\bm v}\in \V_d\right] | {\bm v}\in \mathcal{N}_j\right\}, \quad j=1, \cdots, M.
\end{equation}
Hence, it is feasible to input the occupancy information of the neighbors of $({x_d}_j, {y_d}_j,{z_d}_j)$, denoted as ${\bm O}_{\mathcal{N}_j}$, into the neural network $f(\cdot;{\bm \theta})$ to predict its interpolation pattern $\I_u(j)$:
\begin{equation}\label{eq:simple_model}
    \I_u(j) = f\left({\bm O}_{\mathcal{N}_j};{\bm \theta}\right).
\end{equation}

\SubSection{Overfitting of Lightweight Super Resolution Network}
The design of the network $f(\cdot;{\bm \theta})$ employs a lightweight multi-layer perceptron (MLP) comprising just one hidden layer. 
The input dimension of the network corresponds to the number of elements in the neighborhood set $\mathcal{N}_j$, denoted as $\vert\mathcal{N}_j\vert$.
The number of neurons in the hidden layer is configured as a small positive integer, for instance, $32$, and the output dimension corresponds to the number of elements in the interpolation pattern ${\I_u}(j)$, which is $8$.
To introduce non-linear fitting capability, a Sine activation function~\cite{sitzmann2020implicit} is employed in the hidden layer, while a Sigmoid activation function is applied to the output layer. 
The network $f(\cdot;{\bm \theta})$ contains only a few hundreds or thousands of floating-point numbers, which is significantly smaller compared to networks with millions of parameters.

The prediction of the interpolation pattern can be regarded as eight binary classification problems, and the binary cross-entropy (BCE) loss is used for network overfitting. The BCE loss is formulated as
\begin{equation}
    \ell = \frac{1}{M}\sum_{j=1}^{M} {\I_u}(j)\log f_j+(1-{\I_u}(j))\log (1-f_j),
\end{equation}
where $f_j$ is a simplified representation of $f\left({\bm O}_{\mathcal{N}_j};\btheta\right)$. Batch-wise training is performed using the Adam optimizer~\cite{kingma2015adam} with a learning rate of $0.001$, a batch size of $2048$, and a total training epochs of $150$.

\begin{figure}[!t]
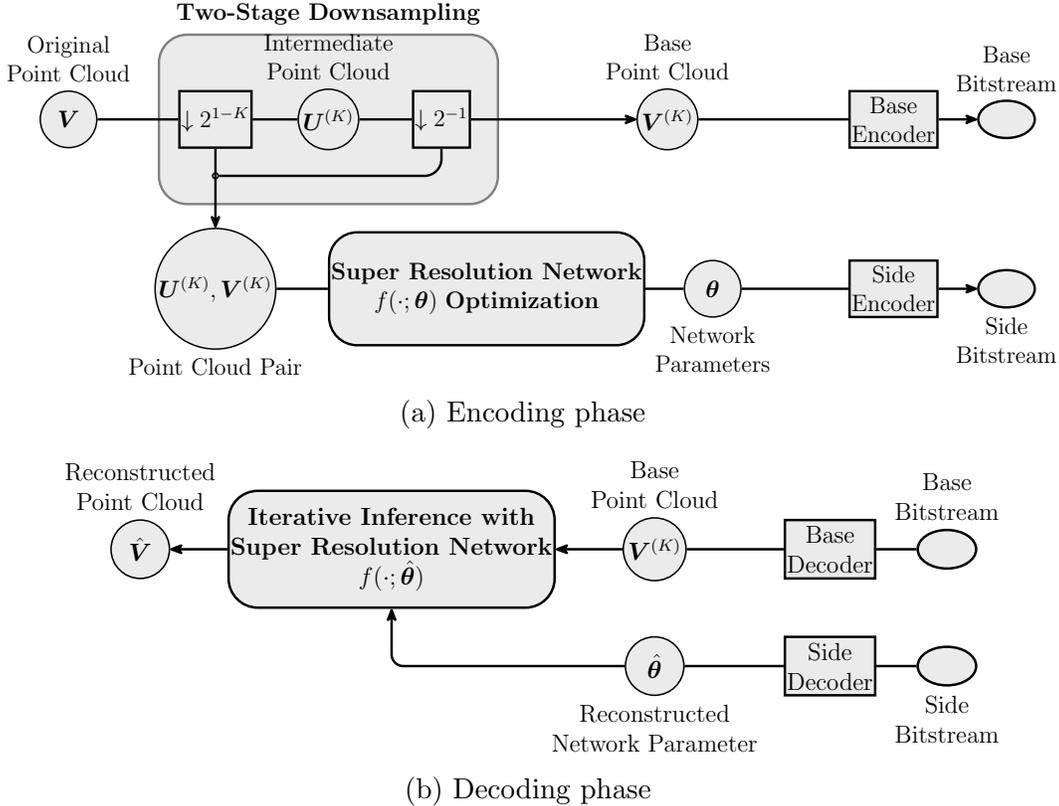

\centering
\resizebox{.95\linewidth}{!}{
\hpsrpccEncoder
}

{\small (a) Encoding phase}
\vspace{3mm}

\resizebox{.85\linewidth}{!}{
\hpsrpccDecoder
}

{\small (b) Decoding phase}
\caption{The proposed point cloud geometry compression framework based on the overfitted lightweight super resolution network.}\label{fig:LSRN-PCGC}
\end{figure}

\SubSection{Overall Framework}
Finally, we propose an innovative point cloud geometry compression approach based on the overfitted lightweight super resolution network. The overall framework of this approach is visualized in Fig.~\ref{fig:LSRN-PCGC}. In the encoding phase, given a point cloud $\V$ and a downsampling factor $q=2^{-K}$, where $K\in\mathbb{N}_+$, we perform a two-step downsampling to obtain the intermediate and base point clouds as follows:
\begin{align}
    \U^{(K)} &=\mathrm{unique}\left(\left[\V/2^{K-1}\right]\right),\\
    \V^{(K)}&=\mathrm{unique}\left(\left[{\bm U}^{(K)}/2\right]\right).
\end{align}
Subsequently, $\U^{(K)}$ and $\V^{(K)}$ form the pair for overfitting the super resolution network $f(\cdot;\btheta)$. Then, the network parameter $\btheta$ and the base point cloud $\V^{(K)}$ are encoded separately. G-PCC~\cite{tmc13v22} or OctAttention~\cite{fu2022octattention} is adopted as the base encoder to losslessly compress $\V^{(K)}$, while the floating-point compression tool fpzip~\cite{lindstrom2006fast} is adopted as the side encoder to compress $\btheta$ with a precision of $16$ bits. In the decoding phase, the losslessly decoded $\V^{(K)}$ is enhanced with the assistance of the reconstructed super resolution network $f(\cdot;\htheta)$. If the interpolated point cloud has not yet reached the scale of the original point cloud, we apply $f(\cdot;\htheta)$ once more to enhance the reconstruction quality further. Finally, the interpolated point cloud is upscaled to match the scale of $\V$.

\Section{Experiments}
The proposed method, named LSRN-PCGC, is evaluated on MPEG Cat1 (Solid) and Cat2 datasets~\cite{gpcc-ctc,mpeg2020vpcc-ctc}. Six rate points are considered corresponding to $ K=1,2,3,4,5,6$, respectively. For Cat1 (Solid), we consider $26$ neighbors where the distance along each axis does not exceed $D=1$ units as the neighborhood set $\mathcal{N}_j$ and set the hidden size of the network to $2^{6-K}$ correspondingly for balancing the bitrate cost of the base point cloud and the network parameter. For Cat2,  we employ a frame sampling rate of $10$ to expedite training, while testing on all frames. Since the base bitstream is relatively large, for Cat2, we consider $124$ neighbors whose distances to the current point along each axis do not exceed $D=2$ units, and the hidden size of the network is kept constant at $32$. We use point-to-point (D1) distance as the distortion measurement~\cite{mekuria2017design} and report the Bj{\o}ntegaard-delta bitrate (BDBR)~\cite{VCEG-M33} to evaluate rate-distortion performance gains. For those interested in the implementation, it is available at the following URL: \url{https://github.com/lidq92/LSRN-PCGC}.

\SubSection{Results on MPEG Cat1 (Solid)}
In Table~\ref{tab:Cat1solid_bdbr}, we provide a comparison of LSRN-PCGC with G-PCC (octree) v22~\cite{tmc13v22} and the lookup table-based method HPSR-PCGC~\cite{li2023hierarchical}.
Compared to G-PCC, LSRN-PCGC exhibits significant D1-BDBR savings, ranging from $65.4\%$ to $80.4\%$, with an average of $74.5\%$ on the MPEG Cat1 (Solid) dataset. 
In comparison to HPSR-PCGC~\cite{li2023hierarchical}, LSRN-PCGC achieves an average of $20.2\%$ D1-BDBR savings. This can be attributed to its capability to access more neighbors without incurring an exponential increase in bitrate cost.
These results verify the effectiveness of the lightweight super resolution network in improving rate-distortion performance.

\begin{table}[!t]
\centering
\caption{D1-BDBR savings for LSRN-PCGC against G-PCC v22~\cite{tmc13v22} and HPSR-PCGC~\cite{li2023hierarchical}} \label{tab:Cat1solid_bdbr}
\vspace{2mm}
\begin{tabular}{lcc}
  \toprule
  \multirow{2}{*}{Point Cloud} & LSRN-PCGC & LSRN-PCGC \\
  &  vs. G-PCC &  vs. HPSR-PCGC  \\
  \midrule
  basketball\_player\_vox11\_00000200 & $-75.0\%$ & $-14.0\%$  \\
  dancer\_vox11\_00000001 & $-74.7\%$ & $-18.3\%$ \\
  facade\_00064\_vox11 & $-80.4\%$ & $-26.8\%$ \\
  longdress\_vox10\_1300 & $-74.9\%$ & $-25.7\%$ \\
  loot\_vox10\_1200 & $-74.0\%$ & $-17.1\%$ \\
  queen\_0200 & $-76.9\%$ & $-23.4\%$ \\
  redandblack\_vox10\_1550 & $-73.2\%$ & $-24.0\%$ \\
  soldier\_vox10\_0690 & $-75.8\%$ & $-26.9\%$ \\
  thaidancer\_viewdep\_vox12 & $-65.4\%$ & $-5.5\%$ \\
  \midrule
  \textbf{MPEG Cat1 (Solid) Average} & $-74.5\%$ & $-20.2\%$ \\
  \bottomrule
\end{tabular}
\end{table}

\begin{figure}[!t]
    \centering
    \includegraphics[width=.6\linewidth]{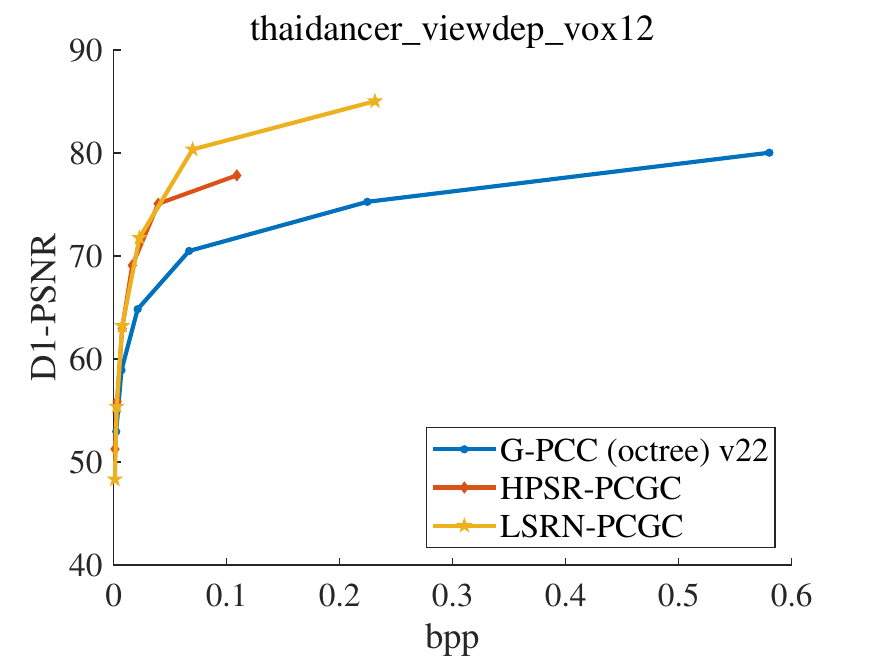}
    \caption{Rate-distortion curves comparison between LSRN-PCGC, HPSR-PCGC, and G-PCC (octree). The horizontal axis represents the rate measurement (bits per point, bpp) for geometry, and the vertical axis represents the distortion measurement (D1-PSNR).}
    \label{fig:Cat1solid_rd}
\end{figure}

In Fig.~\ref{fig:Cat1solid_rd}, we closely examine the rate-distortion curves of the point cloud with the least D1-BDBR savings (thaidancer\_viewdep\_vox12).
These curves vividly illustrate that both LSRN-PCGC and HPSR-PCGC outperform G-PCC, and the performance gap is substantial. This observation underscores the effectiveness of the super-resolution methods in enhancing the rate-distortion performance.
Notably, for the two higher rate points, where the network parameter consumes only a tiny portion of the total bitstream, the lightweight super-resolution network can achieve an improvement of up to $10$ dB D1-PSNR over G-PCC while incurring smaller bitrate costs, leading to a better rate-distortion performance than HPSR-PCGC.

\SubSection{Results on MPEG Cat2}
In Table~\ref{tab:Cat2_bdbr}, we present a comparison between our method and V-PCC v22~\cite{tmc2v22}. With the aid of the learned lightweight super resolution network, our method utilizing G-PCC (octree) as the base encoder outperforms V-PCC, achieving an average D1-BDBR savings of $38.7\%$ D1-BDBR savings. When using the more advanced base encoder OctAttention~\cite{fu2022octattention}, our method maintains the same D1-PSNR performance but significantly reduces the bitrate cost for encoding the base point cloud. As a result, it saves more than $50\%$ D1-BDBR compared to V-PCC on the MPEG Cat2 dataset. Additionally, when evaluating the four point cloud sequences (loot, redandblack, soldier, and longdress) from 8iVFB, we observe average D1-BDBR savings of $40.1\%$ and $56.2\%$ for LSRN-PCGC with G-PCC (octree) and OctAttention, respectively. This indicates that LSRN-PCGC, utilizing a lightweight neural network with G-PCC (octree), marginally outperforms the reported performance ($39.4\%$) of PCGCv2~\cite{wang2021multiscale} which relies on a large neural network. Furthermore, LSRN-PCGC can significantly outperform PCGCv2 when paired with an advanced base encoder like OctAttention. These results confirm the effectiveness of our method in handling dynamic solid point cloud sequences.

\begin{table}[!t]
\centering
\caption{D1-BDBR savings of LSRN-PCGC against V-PCC v22~\cite{tmc2v22}}\label{tab:Cat2_bdbr}
\begin{tabular}{lccc}
\toprule
  \multirow{3}{*}{Point Cloud Sequence} & {LSRN-PCGC} & {LSRN-PCGC} \\
 & with G-PCC & with OctAttention \\ 
  & {vs. V-PCC} & {vs. V-PCC} \\
\midrule

loot & $-33.8\%$ & $-52.2\%$ \\
redandblack & $-46.2\%$ & $-61.3\%$ \\
soldier & $-45.0\%$ & $-58.7\%$ \\
queen & $-44.6\%$ & $-59.0\%$ \\
longdress & $-35.3\%$ & $-52.7\%$ \\
basketball\_player\_vox11 & $-31.0\%$ & $-42.5\%$ \\
dancer\_player\_vox11 & $-35.1\%$ & $-46.7\%$ \\
\midrule
\textbf{8iVFB Average} & {$-40.1\%$} & {$-56.2\%$}\\
\midrule
\textbf{MPEG Cat2 Average} & {$-38.7\%$} & {$-53.3\%$}\\
\bottomrule
\end{tabular}
\end{table}

\Section{Discussion and Conclusion}
We have introduced a content-adaptive approach utilizing a lightweight super-resolution network for lossy point cloud geometry compression. Our experiments on the MPEG Cat1 (Solid) and Cat2 datasets have validated the effectiveness of our proposed method in enhancing rate-distortion performance and its superiority over lookup table-based methods. While this work represents a preliminary stage, it can be extended in several directions. First, incorporating attribute information can enhance the accuracy of predicted interpolation patterns, further improving compression performance. Second, although our method employs a lightweight decoding process, network overfitting increases the encoding time. To address this, we may explore learning a hypernetwork to reduce encoding time. Third, there is potential to investigate joint lossy/lossless compression of point cloud geometry and attributes within this framework, which can be a promising avenue for further research.

\Section{References}
\bibliographystyle{IEEEbib}
\bibliography{LSRN-PCGC}

\end{document}